%% file: main.tex
\documentclass{article}

\usepackage[utf8]{inputenc}
\usepackage{graphicx} 
\usepackage{mathtools}
\usepackage{xltabular}
\usepackage{amsmath}
\usepackage{upgreek}
\usepackage{hyperref}
\usepackage{ragged2e}
\usepackage[a4paper, top=2.5cm, bottom=2.5cm, left=2.5cm, right=2.5cm]{geometry}
\date{}   
\input{commands}

\begin{document}

\begin{center}
    {\LARGE Tunable Dynamic Speckle Generation for Random Illumination Microscopy \par}
    \vspace{1em}
    { Lilian Magermans\textsuperscript{1}, Assia Benachir\textsuperscript{1}, Nathan P. Spiller\textsuperscript{2}, Tianxin Wang\textsuperscript{2}, Federico Vernuccio\textsuperscript{1}, Randy Bartels\textsuperscript{3}, Stephen M. Morris\textsuperscript{2}, Steve J. Elston\textsuperscript{2}, Martin J. Booth\textsuperscript{2}, Hervé Rigneault\textsuperscript{1}*\par}
    \vspace{0.5em}
    {\textit{\textsuperscript{1}Aix Marseille Université, CNRS, Centrale Med, Institut Fresnel, Marseille, France} \par \textit{\textsuperscript{2}Department of Engineering Science, University of Oxford, Parks Road, Oxford, OX1 3PJ, UK} \par \textit{\textsuperscript{3}Morgridge Institute for Research, Madison, WI, USA}}
    \par
    \vspace{1em}
    *Email address: herve.rigneault@fresnel.fr
\end{center}

\justifying

\begin{abstract}
\input{abstract}
\par
\vspace{0.7em}
\end{abstract}

\section{Introduction}
\input{introduction}

\section{Results}
\input{LC_characterization}
\input{DSI_optical_sectioning_and_speed}
\input{RIM_lateral_resolution}

\section{Conclusion}
\input{Conclusion}

\section{Experimental Section}
\input{LC_preparation}
\input{LC_texture_imaging}

\input{domain_size}
\input{Widefield_experimental}
\input{Decorrelation_time_measurements}
\input{RIM_reconstruction}

\medskip
\textbf{Supporting Information} \par 
Supporting Information is available from the Wiley Online Library or from the author.

\medskip
\textbf{Acknowledgements} \par
We acknowledge S. Massieau and T. Mangeat for providing the U2OS cell and the mouse intestinal villi samples, respectively.
We acknowledge the financial support from the Center National de la Recherche Scientifique (CNRS), A*Midex (Grant No. ANR-11-IDEX-0001-02), ANR (Grant Nos. ANR-10-INSB-04-01, ANR-11-INSB-0006, ANR-16-CONV-0001, and ANR-21-ESRS-0002 IDEC), and INSERM Grant No. 22CP139-00. Chan Zuckerberg Initiative (DAF 2024-337798). This project has received funding from Horizon Europe (Marie Skłodowska-Curie Postdoctoral Fellowship Grant No. 101148683 CHIMERA) and the European Research Council (ERC, SpeckleCARS, Grant No. 101052911).
N.P.S. gratefully acknowledges the Engineering and Physical Sciences Research Council (EPSRC) UK for financial support through a graduate student scholarship (EP/T517811/1).
S.M.M., M. J. B., and S.J.E., gratefully acknowledge financial support from the EPSRC UK through project EP/W022567/1.

\bibliography{biblio_full}

\end{document}

%% file: commands.tex
\usepackage{siunitx} 
\usepackage{xspace}  
\usepackage{caption}
\usepackage{float}  

\newcommand{\Vperum}{\si{\volt\per\micro\meter}\xspace} 
\newcommand{\um}{\si{\micro\meter}\xspace} 
\newcommand{\rbar}{$\bar{r}$\xspace} 
\newcommand{\stdI}{$\langle\sigma(I)\rangle$\xspace} 
\newcommand{\figwidth}{0.85\textwidth}

%% file: abstract.tex
Speckled illumination enhances widefield fluorescence microscopy by enabling optical sectioning and super resolution. In random illumination microscopy, sequences of speckled illumination patterns are used to excite fluorescent samples and images are reconstructed based on a statistical analysis of the intensity fluctuations. Although random illumination microscopy has been shown to give excellent performance, its widespread implementation is hindered by the high cost and complexity of the generation of suitable speckled illumination patterns, which is achieved using digital micro-mirror devices or spatial light modulators. Here, we present a zwitterion-doped liquid crystal (LC) device capable of generating independent, high-contrast speckle patterns with a tunable decorrelation time in the 0.1 s–0.1 ms range under visible laser illumination. This LC-based dynamic speckle generator is applied to widefield random illumination fluorescence microscopy of tissue and cell samples, where it enables optical sectioning with a 2 µm axial resolution, and a 1.5-fold improvement in lateral spatial resolution. Owing to its low cost and simplicity, this LC speckle generator offers an attractive alternative to digital micro-mirror and spatial light modulator devices for implementing widefield random illumination microscopy.

%% file: introduction.tex
Dynamic speckle illumination (DSI) is a widefield microscopy technique in which a sample is illuminated by subsequent speckle patterns in order to obtain optically sectioned images with an axial resolution similar to that of a confocal microscope.\cite{ventalon_quasi-confocal_2005} The enhanced axial resolution originates from the computation of the standard deviation for each pixel based on a stack of many speckled images. Moreover, super-resolution images can be generated through a reconstruction algorithm as is done in random illumination microscopy (RIM).\cite{mudry_structured_2012,mangeat_super-resolved_2021,mazzella_extended-depth_2024}.

Experimentally, such techniques require a reliable generation of statistically independent speckle patterns with a high contrast. Moreover, the speckle generation rate should be tunable as image acquisition parameters, such as the exposure time and frame rate, can vary drastically between (biological) samples depending on their fluorescence intensity. Typically, this is achieved by generating random phase masks which change between images using either a spatial light modulator (SLM) \cite{ventalon_dynamic_2006,mazzella_extended-depth_2024} or a digital micro-mirror device (DMD).\cite{chakrova_studying_2015}  However, SLMs and DMDs are rather expensive and complex, which creates a high barrier for widespread implementation of speckle-based imaging. Another approach is to generate speckle patterns by moving a diffuser with between acquisitions, but this doesn't generate fully randomized patterns which can introduce artifacts.\cite{mudry_structured_2012,yin_dynamic_2024} Moreover, all of these approaches require a synchronization between the camera and the speckle pattern generator, which complicates the experimental setup. To circumvent this, a continuous dynamic speckle generation method could be used, as long as the speckle patterns change at the same timescale as the acquisition rate of the camera. Such a dynamic speckle could be generated by continuously rotating a diffuser, as has been done for speckle reduction applications.\cite{stangner_step-by-step_2017} However, this approach is not well adapted for DSI and RIM, as it is challenging to precisely control the speckle dynamics.

Recently, liquid crystals (LCs) have gained interest as dynamic speckle generators for applications in ghost imaging and laser speckle reduction.\cite{hansford_speckle_2016,hansford_enhancing_2021,jin_laser_2021,davletshin_ghost_2022,jin_zwitterion-doped_2023} When the LC is exposed to an alternating electric field, electrohydrodynamic instabilities are induced, generating a turbulent scattering state. However, precise control of the dynamics of these instabilities and the resulting speckle patterns, which is required for its application in DSI and RIM, has not been demonstrated. 

In this work, we use a zwitterion-doped LC device to generate statistically independent speckle patterns, through dynamic scattering, and demonstrate its application for DSI and RIM. We show that the speckle decorrelation time can be precisely tuned from $10^{-4}$ to $10^{-1}$ seconds through the control of the applied electric field. We implement the LC device in a DSI widefield microscope and achieve an axial resolution of 2 \um. Optically sectioned images are obtained in a mouse intestinal jejunum through DSI. The acquisition speed limit is explored and we report a final DSI frame rate up to 14 Hz, which is compatible with live imaging. Finally, we show that treating the speckled images with the RIM algorithm results in a 1.5 times improvement in lateral resolution compared to the standard uniform illumination widefield imaging.

%% file: LC_characterization.tex
\subsection{Speckle generation using zwitterion-doped nematic LC}
\begin{figure}[H]
\centering
  \includegraphics[width=\figwidth]{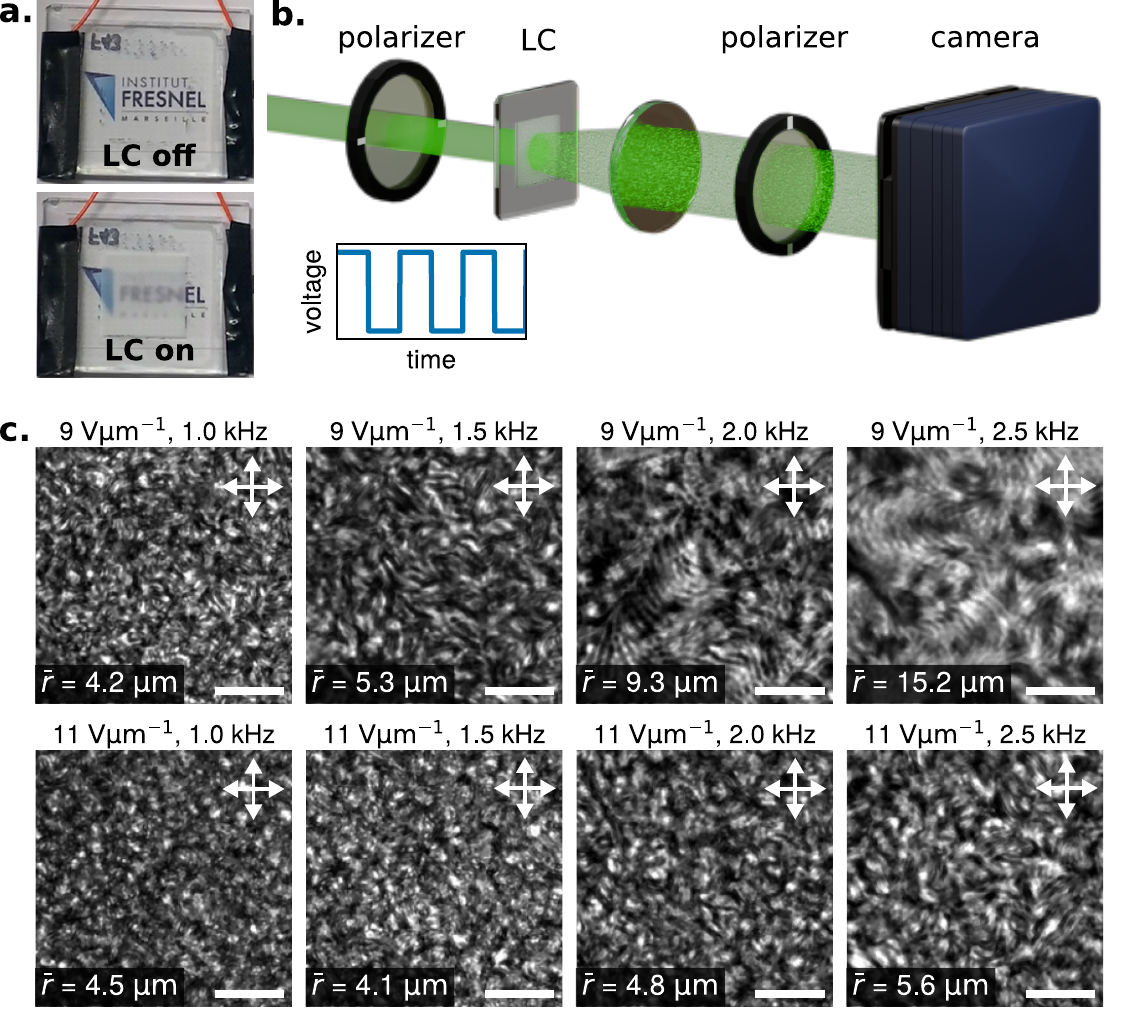}
  \caption{\textbf{Liquid crystal (LC) dynamic speckle generator.} \textbf{a)} Images of the LC device in its OFF (no voltage applied) and ON (voltage applied) state. The active area is 10-by-10 mm. \textbf{b)} Experimental configuration used for imaging LC texture between crossed polarisers whilst applying an electric field across the LC layer. Additional details can be found in the experimental section. \textbf{c)} Images of the LC textures for varying strength and frequency of the alternating electric field. The arrows represent the orientation of the crossed polarisers. The mean domain size \rbar was determined from the computed autocorrelation functions as detailed in the methods. All scale bars represent 25 \um.}
  
  \label{fig:LC textures}
\end{figure}

The LC dynamic speckle generator consisted of a zwitterion-doped nematic LC  filled into a glass cell with an air gap of 20 \um with transparent (indium tin oxide) electrodes coated onto the inner substrates of the glass slides, similar to our previously reported system.\cite{jin_zwitterion-doped_2023} When the device is switched off, the LC align homeotropically (perpendicular to the confining substrates), resulting in a fully transparent appearance when viewed with the naked eye. (\textbf{Figure \ref{fig:LC textures}a}). Applying an alternating electric field to the device results in the formation of electrohydrodynamic instabilities triggered by the competing torques originating from the conductivity and dielectric coupling.\cite{meyerhofer_electrohydrodynamic_1975} These turbulent instabilities produce a dynamic scattering state with a milky appearance (Figure \ref{fig:LC textures}a). \\

The LC optical texture was imaged between crossed polarisers using the setup in \textbf{Figure \ref{fig:LC textures}b} to characterize the dynamic response as a function of the frequency $f$ and field strength $E$ of the square waveform of the applied electric field across the device. Images taken between crossed polarisers show a grainy texture whose mean domain size \rbar strongly depends on the operating conditions (\textbf{Figure \ref{fig:LC textures}c}). This characteristic size is impacted by three main contributions; the dielectric torque aligning the LC director perpendicular to the electric field, the conductive torque generated by the motion of mobile charges and the elastic torque which acts to restore long range orientational order.\cite{dubois-violette_hydrodynamic_1971} Stronger fields produce smaller domains, as both the dielectric and conductive torque are increased and thus overpower the elastic torque. Conversely, an increase in $f$ typically result in larger domains, which can be understood from the decreased conductive torque. In fact, theory predicts that when $f$ surpasses a cross-over frequency $f_c$ the system enters a purely ``dielectric regime", where the mobile charges can no longer react to the external electric field fluctuations.\cite{dubois-violette_hydrodynamic_1971} In this case the conduction torque is negligible and the system is dominated by the dielectric torque aligning the LC director perpendicularly to the field coupled with the elastic energy favouring long range order, hence why \rbar increases with  $f$. This dielectric regime gives rise to chevron textures and dielectric rolls as observed for $E$ = 9 \Vperum and $f>$1.5 kHz (Figure \ref{fig:LC textures}c).\cite{eber_electrically_2016} The cross-over frequency between the conductive and dielectric regimes increases with an increasing applied field. Therefore, we do not see any dielectric rolls or chevrons at $E$ = 11 \Vperum for any of the frequencies probed. 
It should also be noted that the variation in domain size strongly impacts the divergence of the light beam exiting the LC. Similarly to a one-dimensional diffraction grating, the effective numerical aperture scales inversely with \rbar (\textbf{Figure S1}).
\\

\begin{figure}[H]
\centering
  \includegraphics[width=\figwidth]{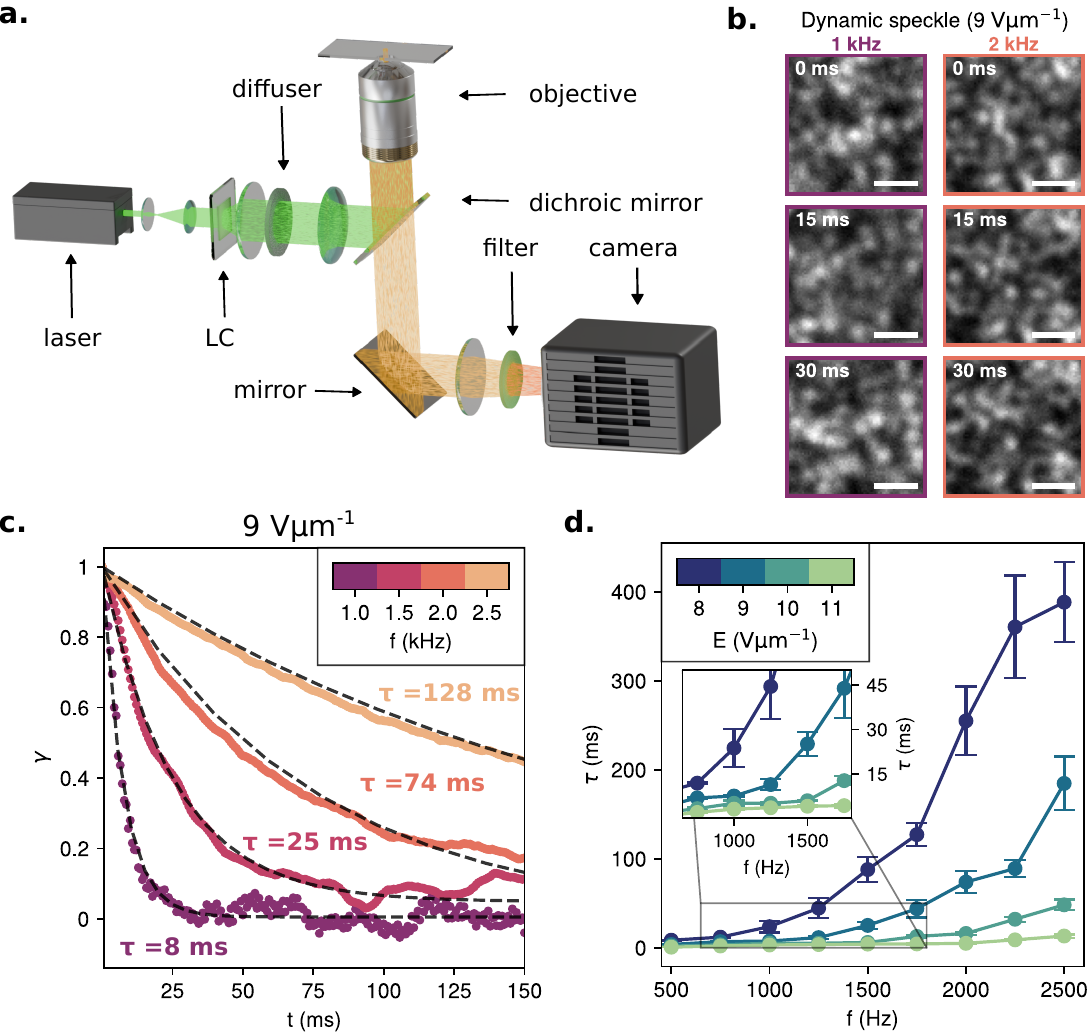}
  \caption{\textbf{Dynamics of speckled illumination for widefield microscopy.} \textbf{a)} Experimental setup. Speckled illumination is generated by a continuous laser ($\uplambda$=532 nm) which passes through the LC device followed by a thin diffuser which eliminates the zero-order. The LC is imaged on the back focal plane of the objective, generating a widefield speckled illumination. Fluorescence intensity is collected in the epi-configuration and captured by a scientific camera. \textbf{b)} Zoomed images of a thin, uniform fluorescent film under dynamic speckle illumination whilst operating the LC at 9 \Vperum for two different frequencies over time. Scale bars are 2 \um. \textbf{c)} Decorrelation curves of speckle patterns at 9 \Vperum for various values of $f$. The correlation coefficient $\gamma$ is computed between the first image and an image taken at time $t$. Black dashed lines indicate an exponential decay fit whose characteristic decay time is the decorrelation time $\tau$. \textbf{d)} Decorrelation time $\tau$ as a function of the applied electric field frequency and amplitude.}  
  
  \label{fig:widefield_dynamics}
\end{figure}

Next, we implement the LC device in the microscopy setup and investigate the dynamics of the speckle patterns generated. \textbf{Figure \ref{fig:widefield_dynamics}a} shows the optical configuration of the widefield microscope. The excitation beam is expanded to fit the active area of the LC and a diffuser is used to minimize any zero-order (non-diffracted light) from the speckled beam. To generate a widefield speckle illumination, the LC is imaged onto the back focal plane of the objective. The collected fluorescence signal is captured with a scientific camera capable of acquiring images up to 1400 frames per second. The evolution of the speckle patterns is recorded by imaging a glass cover slip coated with a thin fluorescent layer. \textbf{Figure \ref{fig:widefield_dynamics}b} shows snapshots at three different timestamps for different operating conditions. The size of the speckle grains remains constant under all operating conditions, as it depends on the numerical aperture of the objective and is not impacted by the domain size of the LC texture. Whereas all images are drastically different for $f =$ 1 kHz, changes in the speckle pattern are more gradual for $f = $ 2 kHz, indicating a difference in dynamics. To quantify this, we compute the correlation coefficient $\gamma$ between the first image and the image at time $t$ as described in the experimental section. The evolution of $\gamma(t)$ follows an exponential decay curve whose time constant describes the decorrelation time $\tau$ (\textbf{Figure \ref{fig:widefield_dynamics}c}). Indeed, we find that the speckle patterns change almost ten times faster for $f$ = 1 kHz than for $f$ = 2 kHz when $E$ = 9 \Vperum. This is in line with the observation that \rbar is smaller for lower $f$ (Figure \ref{fig:LC textures}c), as small domains are a sign of increased turbulence and can change orientation more rapidly. Indeed, increasing $E$ also generates faster decorrelation across all frequencies as expected from the LC texture observations (\textbf{Figure \ref{fig:widefield_dynamics}d}). Finally, by carefully controlling the parameters of the alternating electric field, the decorrelation time $\tau$ can be tuned from sub-millisecond to hundreds of milliseconds. 

%% file: DSI_optical_sectioning_and_speed.tex
\subsection{Dynamic speckle illumination (DSI) microscopy using LC}
\begin{figure}[H]
\centering
  \includegraphics[width=\figwidth]{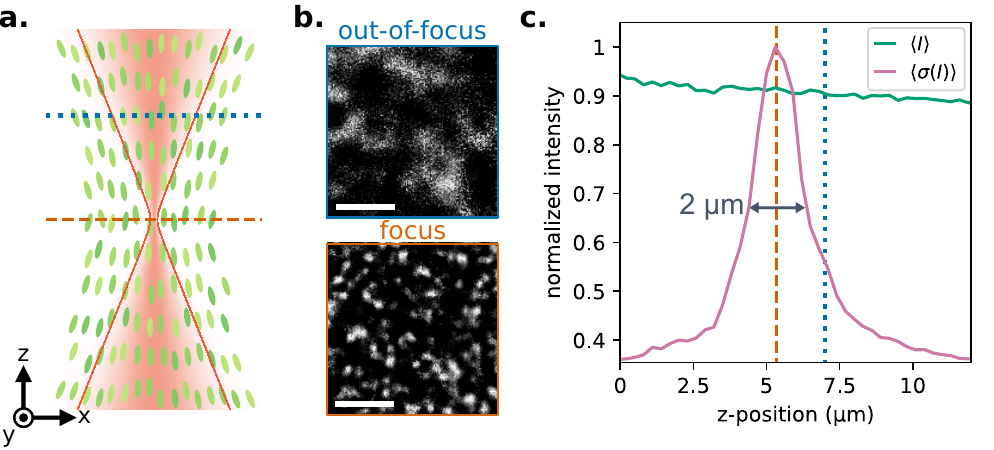}
  \caption{\textbf{Optical sectioning by dynamic speckle imaging (DSI)} \textbf{a)} Schematic illustration of the excitation speckle field in green and the detection point spread function (PSF) in red. The two dashed lines indicate the focal plane (orange) and an out-of-focus plane (blue). \textbf{b)} Normalized images of speckle generated on a fluorescent thin film of the out-of-focus and in-focus planes. Scale bars are 5 \um. \textbf{c)} Demonstration of optical sectioning by acquiring image stacks at different z-positions. For each stack, we compute the average intensity image $\bar{I}$ and the standard deviation image $\sigma(I)$ and plot their spatial averages (indicated by $\langle$ $\rangle$) for each z-position. The full width half maximum (FWHM) of \stdI is indicated in grey.}  
  
  \label{fig:optical_sectioning_highlighter}
\end{figure}

As an application of the LC dynamic speckle generator, we first investigate its use for DSI imaging. One of the main advantages of DSI over standard widefield imaging is the optical sectioning arising from the statistical analysis. \textbf{Figure \ref{fig:optical_sectioning_highlighter}a} illustrates the widefield speckle illumination intensity used for DSI as well as the point spread function (PSF) of the objective. Whereas the speckle grains are mostly uniform throughout the depth of the sample, the detection PSF has a narrow waist only at the focal plane. This means that a fluorescence source located outside the focal plane appears larger and blurred, which is clearly visible when imaging a uniform fluorescent film at different z-positions (\textbf{Figure \ref{fig:optical_sectioning_highlighter}b}). In the out-of-focus plane, the intensity of a single pixel within the image will only vary slightly when the speckle illumination pattern changes due to this spatial blur, whereas the intensity fluctuates significantly close to the focal-plane. To quantify this, we compute the average intensity image $\bar{I}$ and standard deviation image $\sigma(I)$ for image stacks acquired at different z-positions. Taking the spatial average, indicated by $\langle$ $\rangle$, of these images demonstrates that even though  $\langle\bar{I}\rangle$ remains relatively constant, $\langle\sigma(I)\rangle$ changes drastically and is much higher near the focal plane.  (\textbf{Figure \ref{fig:optical_sectioning_highlighter}c}), providing the z-sectioning capability of DSI. \cite{ventalon_quasi-confocal_2005} By considering \stdI measured whilst moving the z-position of a thin fluorescent sample, the axial resolution of the system is found to be 2 \um (Figure \ref{fig:optical_sectioning_highlighter}c).

\begin{figure}[H]
\centering
  \includegraphics[width=\figwidth]{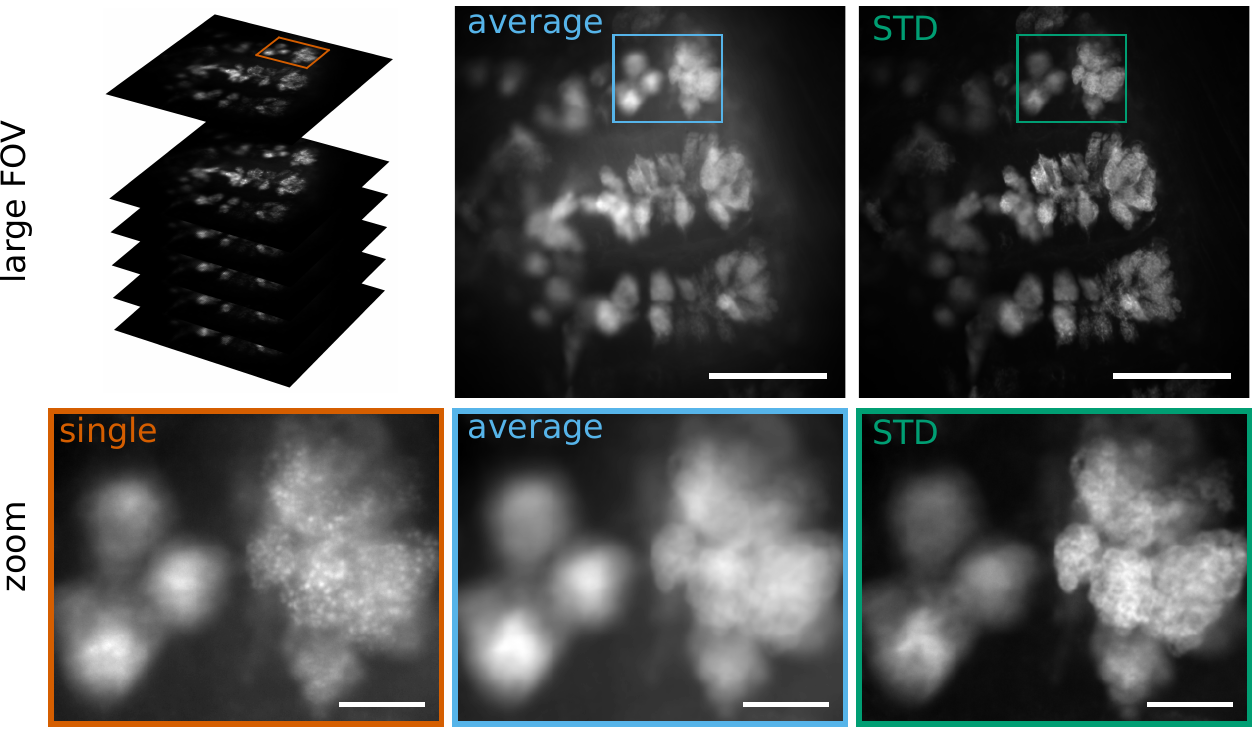}
  \caption{\textbf{Demonstration of optical sectioning using DSI in a mouse intestinal jejunum labeled with rhodamine.} A stack of images with different speckle illumination patterns is acquired (left column) with an exposure time $t_{exp} =$ 50 ms, whilst operating the LC at $\tau \approx$ 55 ms. A conventional widefield image is reconstructed by computing the average intensity $\bar{I}$ image of the stack (middle column). The standard deviation $\sigma(I)$ provides an optically sectioned image with improved contrast and resolution (right column). Scale bars are 50 \um for the large FOV and 20 \um for the zoomed images.}  
  
  \label{fig:mouse_intestines}
\end{figure}

Another advantage of DSI is that speckled illumination is not disturbed by scattering and aberrations, making it well suited for biological samples.\cite{goodman_speckle_2020} To demonstrate this, we image rhodamine-marked intestines of a mouse which have three-dimensional features spanning tens of \um in depth (\textbf{Figure \ref{fig:mouse_intestines}}). To get well contrasted image frames the LC dynamics were tuned such that the speckle patterns decorrelated slightly slower than the camera exposure time ($t_{exp} =$ 50 ms and $\tau \approx$ 55 ms). Individual speckle images show well defined speckle grains only where the fluorophores are within the focal plane (Figure \ref{fig:mouse_intestines} left column). This high speckle contrast indicates that the dynamics of the LC device are well tuned to produce single speckle patterns rather than multiple patterns blurred together. We can clearly observe the optical sectioning achieved by DSI by comparing the standard deviation $\sigma(I)$ image to the average intensity $\bar{I}$ image (Figure \ref{fig:mouse_intestines} middle and right column). The $\bar{I}$ image, equivalent to the standard uniform illumination image, shows a more uniform intensity across the sample, whereas in the $\sigma(I)$ image the intensity of out-of-focus regions is significantly reduced. Moreover, DSI noticeably improves the lateral resolution as the fine features in the sample become distinguishable.  

\begin{figure}
\centering
  \includegraphics[width=\figwidth]{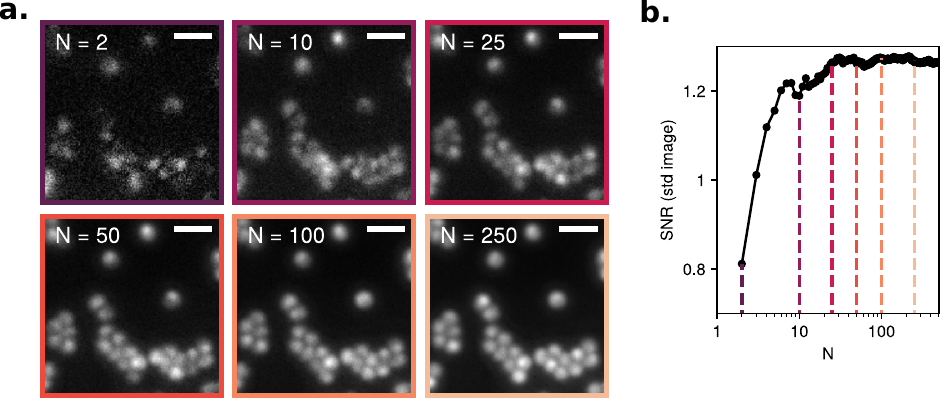}
  \caption{\textbf{High speed DSI imaging of fluorescent beads.} Images are acquired at 1400 FPS, with $t_{exp} =$ 700 \um and $\tau \approx$ 2 ms. \textbf{a)} Standard deviation $\sigma(I)$ images based on a stacks with a different number frames $N$. Scale bars are 2.5 \um. \textbf{b)} Signal to noise ratio (SNR) of the generated $\sigma(I)$ image based on $N$ frames. The SNR is approximated as the ratio between the mean value and spatial standard deviation of each $\sigma(I)$ image. Dashed lines correspond to the images in a).}  
  
  \label{fig:beads}
\end{figure}

A limitation of DSI is that it requires a stack of many images to obtain a single optically sectioned, well contrasted image. The effective frame rate therefore does not only depend on the exposure time of the camera but also on the total number of images required to obtain the $\sigma(I)$ image. To optimise this frame rate, the LC dynamics should be carefully tuned. The speckle pattern should decorrelate fast enough to generate distinguishable patterns between frames, without changing so fast that multiple speckle patterns are captured in one frame, resulting in blur. To investigate the limit of the effective frame rate achievable with our setup, we acquire images of brightly fluorescent beads with the minimum exposure time of our camera (700 \um). \textbf{Figure \ref{fig:beads}a} shows the reconstructed standard deviation $\sigma(I)$ images for different numbers of frames $N$. To get a good $\sigma(I)$ image, each pixel in the FOV should be illuminated by a bright speckle grain several times. At low $N$ ($<$ 25), not all beads are visible in the $\sigma(I)$ image, which indicates that only part of the FOV has been illuminated with bright speckle grains. When more frames are considered ($25<N<50$), all features become visible but the boundaries between beads remain blurry due to a lack of statistics on the intensity fluctuations. For $N\approx$ 100, we start to see the improved lateral resolution of DSI in the well defined boundaries between beads. Moreover, considering 100 frames seems to be sufficient for obtaining optical sectioning in three-dimensional samples (\textbf{Figure S2}). Further increasing $N$ improves the quality of the image, as can be clearly observed when plotting the signal to noise ratio (SNR) as a function of $N$ (\textbf{Figure \ref{fig:beads}b}). 
There is always a trade-off between the effective frame rate and the SNR, but generally $100$ frames is enough to reconstruct an optically sectioned image capturing all features as demonstrated in Figure \ref{fig:beads}a and on different samples in the SI (\textbf{Figures S2-4}). Using the current setup, it should therefore be possible to acquire optically sectioned images at 14 Hz (100 images acquired in 70 ms giving one DSI image). Moreover, the current limitation is the camera's acquisition speed rather than the decorrelation time of the LC dynamic speckle generator. It should therefore be possible to further decrease the acquisition time if a high speed camera is used to match the sub-\um decorrelation time achievable with the LC device. This opens up the possibility replace the SLM to perform optically sectioned live widefield imaging at a low-cost, which is of large interest for biological applications. 

%% file: RIM_lateral_resolution.tex
\subsection{Improved lateral resolution through Random Illumination Microscopy (RIM)}
\begin{figure}[H]
    \centering
    \includegraphics[width=\figwidth]{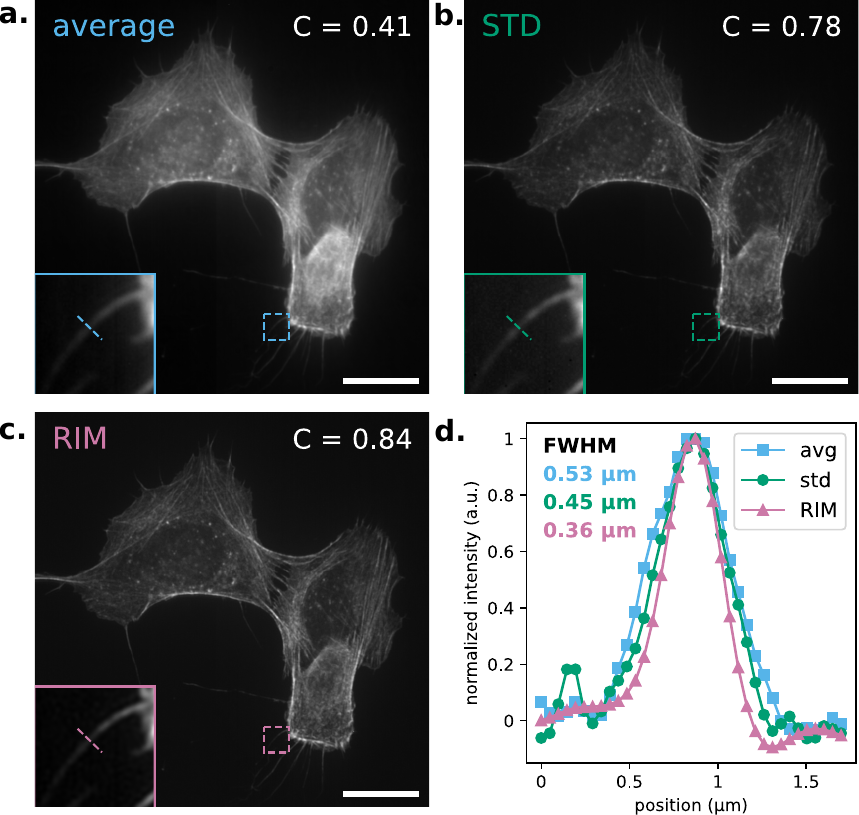}
    \caption{\textbf{Enhanced lateral resolution using RIM algorithm demonstrated on U2OS cells with labelled actin (Alexa Fluor 546 phalloidin)}. Images are acquired at 10 FPS with $t_{exp} = $ 100 ms and $\tau \approx$ 100 ms. \textbf{a-c)} Computed average, standard deviation and RIM images respectively, based on an image stack of speckled images. The contrast $C$ is computed as the spatial standard deviation divided by the mean intensity of each image. Insets are a zoom of the region indicated by the dashed boxes and show a thin bundle of actin filaments. All scale bars represent 20 \um. \textbf{d)} Normalized intensity profile collected across the dashed line in the zoomed regions in a-c) for each type of image. The full width half maximum (FWHM) is indicated in colours matching the legend.} 
    \label{fig: RIM}
\end{figure}

Further enhancement to the spatial resolution and contrast can be achieved by processing the stacks of speckled images with the RIM algorithm. This algorithm retrieves the fluorescence intensity distribution by iterative optimization of a theoretical model to match the experimentally obtained $\sigma(I)$.\cite{mangeat_super-resolved_2021,affannoukoue_super-resolved_2023}. Here, we demonstrate its impact by imaging a sample of U2OS cells whose actin has been labelled with a fluorophore. We acquire speckled images using the LC device and compute the average intensity, standard deviation and RIM images (\textbf{Figure \ref{fig: RIM}}). We find a two-fold improvement in the optical contrast when comparing the average (uniform illumination) to the computed RIM image. To evaluate the change in lateral resolution, the intensity profile across a narrow bundle of actin filaments is plotted, as indicated by the dashed line in the insets of Figure \ref{fig: RIM}. As previously observed (\textbf{Figure \ref{fig:mouse_intestines}}), the lateral resolution is already enhanced by computing $\sigma(I)$ as compared to $\bar{I}$ by a factor of 1.18 (Figure \ref{fig: RIM}d). However, employing the RIM algorithm improves the lateral resolution more significantly, showing a 1.5 times improvement compared to $\bar{I}$ (\textbf{Figure \ref{fig: RIM}d}). 

%% file: conclusion.tex
 In this study, we demonstrated the use of a zwitterion-doped LC dynamic speckle generator for the application of speckle imaging techniques such as DSI and RIM. Its compact design and straightforward operation would facilitate a simple integration into a commercial wide-field microscope to enable speckled illumination imaging. Characterization of the LC device shows that the speckle decorrelation time can be precisely controlled from sub-millisecond up to hundreds of milliseconds. The LC dynamic speckle generator is implemented in a widefield microscopy setup and optically sectioned images of biological samples are reconstructed using DSI. Based on high speed images acquired, live-imaging should be achievable with the different speckle patterns generated by the LC. Additionally, the application of the RIM algorithm shows a 1.5 times improvement in lateral resolution and a two-fold enhancement in the optical contrast, demonstrating the high quality speckle generated by the LC device. To conclude, the wide tunability of speckle decorrelation time, combined with its low cost and simplicity, make the LC device and excellent match for DSI and RIM and provides a major advancement for the widespread implementation of speckled illumination imaging.

%% file: LC_preparation.tex
\subsection{LC device fabrication}
The LC dynamic scattering devices were assembled using commercially available INSTEC LC cells (S100A200uT180), corresponding to nominal cell gaps of \SI{20 \pm 2}{\micro\metre}. The size of the air gap was defined by \SI{20}{\micro\metre} glass spacer beads distributed throughout the glass cell. Before filling, the cell gaps were confirmed via an Agilent Cary 8454 UV-Vis spectrophotometer by analyzing interference signals across the \SIrange{300}{1000}{\nano\metre} wavelength range. Only cells with a nominal cell gap within a \SI{5}{\percent} error of the quoted \SI{20}{\micro\metre} were used to form the devices for this study. Each cell contained indium-tin-oxide (ITO) electrodes (\SI{23}{\nano\metre} thick) on the inner surfaces of the top and bottom substrates, forming a \SI{10}{\milli\metre} $\times$ \SI{10}{\milli\metre} active region for electric field application. The ITO was coated with a DPI-V011 alignment layer providing a homeotropic LC alignment at the boundaries. The LC mixtures were capillary-filled into the cells at \SI{5}{\degreeCelsius} above the temperature where they phase change to an isotropic liquid (clearing point), to reduce the risk of air bubbles forming inside the device during the process. After fabrication, uniform dispersion of the LC mixture in the cell was confirmed under a polarizing optical microscope (Olympus BX51-P). 

%% file: LC_texture_imaging.tex
\subsection{Imaging of LC texture}
The turbulent flow regime associated with the dynamic scattering mode involves rapid LC director fluctuations occurring on microsecond to millisecond timescales. To resolve some of these dynamics, a custom high-speed polarising optical microscopy system was constructed, as illustrated by the schematic in (\textbf{Figure \ref{fig:LC textures}b}). The LC device is illuminated by a LED (centre wavelength \SI{600}{\nano\metre}) whose output if focused by a condensing lens (f = \SI{30}{\milli\metre}) to a concentrated spot within the field of view. Whilst this configuration sacrifices illumination uniformity compared to Köhler illumination, it was necessary to achieve a sufficient signal-to-noise ratio (SNR) at exposure times below \SI{100}{\micro\second}.\\

The camera used was a Photron FASTCAM Nova S6, which enables capture rates of 6,400 frames per second at full resolution (1024 × 1024 pixels), with the effective exposure time equal to the frame period (\SI{156}{\micro\second} at 6,400 fps). By selecting reduced regions of interest, frame rates can be increased substantially and achieve 16,000 fps at 640 x 640 pixels (effective period = \SI{62.5}{\micro\second}). The trade-off between temporal resolution and spatial field of view was optimised for each experiment. The images displayed in (\textbf{Figure \ref{fig:LC textures}c}) employed 640 x 640 pixel resolution at 16,000 fps, providing \SI{62.5}{\micro\second} \SI{62.5}{\micro\second} temporal resolution over an ~ \SI{250}{\micro\metre} x \SI{250}{\micro\metre} field of view, sufficient to resolve individual domains whilst capturing multiple periods.

%% file: domain_size.tex
\subsection{Measuring LC domain size}
When applying a square electric field, turbulence is induced in the LC device due to electrohydrodynamic instabilities. The captured still images between crossed polarisers (Figure \ref{fig:LC textures}c) show textures whose domain size depends on the field parameters. To quantify the domain size $\bar{r}$, we consider the autocorrelation matrix $G(k,l)$ defined as:
\begin{equation}
    G(k,l) = \sum_{m=0}^{M-1}\sum_{n=0}^{N-1}I(m,n)\bar{I}(m-k,n-l)
    \quad
    \left\{
    \begin{array}{l}
      -(M-1)<k<M-1 \\
      -(N-1)<l<N-1 
    \end{array}
    \right.
    \label{eq:correlation coefficient}
\end{equation}
where $k$ and $l$ correspond to translations of the rows and columns respectively, $I$ is the image, with dimensions $M$-by-$N$, and $\bar{I}$ its complex conjugate. The resulting matrix $G(k,l)$ shows a sharp, radially symmetric peak centred at the origin ($k=0,l=0$). The domain size $\bar{r}$ is approximated as the full width half maximum (FWHM) of this peak. 

%% file: Widefield_experimental.tex
\subsection{Experimental widefield microscopy setup}
Figure \ref{fig:widefield_dynamics}a represents the optical setup used for speckled illumination imaging. The excitation source is a continuous solid state laser ($\lambda_{exc}=532$ nm, LCX-532S-300, Oxxius). The excitation beam is expanded using a telescope configuration ($f_1 =$ 75 mm, $f_2=$ 500 mm) to match the active area of the LC device of 10 mm $\times$ 10 mm. A signal generator (SDG6022X, Siglent) and amplifier (10$\times$ per port, F10AD, Pendulum) are used to control the LC dynamics by applying square waveforms with a controlled potential and frequency. Dynamic speckle patterns are generated by passing the laser through the LC and to minimize the impact of non-scattered components a static diffuser (48-514, Edmund Optics) is added to the beam path. The LC is imaged on the back focal plane of the objective (40$\times$ Plan Achromat Objective, 0.65 NA, Olympus) using a telescope ($f_1 =$ 200 mm, $f_2 =$ 250 mm). For z-scanning a 1-axis motorized stage (L-836.501212, PI) is used to move the objective. Fluorescence is epi-collected and imaged on a scientific camera (ORCA-fusion, C14440-20UP Hamamatsu) using a tube lens ($f=300$~mm) resulting in a final magnification of $\approx 67 \times$. 

%% file: Decorrelation_time_measurements.tex
\subsection{Decorrelation time measurements}
The two-dimensional correlation coefficient $\gamma$ of two matrices $X$ and $Y$ which each have dimensions $M$-by-$N$ is given by:

\begin{equation}
    \gamma = \frac{1}{M\cdot N}\sum_{m=0}^{M-1}\sum_{n=0}^{N-1}X(m,n)\bar{Y}(m,n)
    \label{eq:correlation coefficient_2}
\end{equation}
where $\bar{Y}$ is the complex conjugate of $Y$. To quantify the decorrelation of the speckle patterns over time, we compute $\gamma(t)$ such that $X$ is the first image at time $t_0$  and $Y$ is the image at time $t_0+t$. Each images is normalized by subtracting the mean intensity and dividing by its standard deviation. The decorrelation time $\tau$ is found by fitting $\gamma(t)$ to an exponential decay curve with the form:
\begin{equation}
    \gamma(t) = (1-\gamma_\infty)e^{-t/\tau} +\gamma_\infty
    \label{eq: exponential fit}
\end{equation}
where $\gamma_\infty$ represents the baseline which when it is non-zero indicates the presence of a static speckle pattern. Typically, $\gamma_\infty$ is found to be below 0.05, which indicates a complete decorrelation and thus the generation of statistically independent patterns. 
\\

%% file: RIM_reconstruction.tex
\subsection{RIM reconstruction}
The RIM images are constructed based on 500 speckled images through a Matlab script using the previously reported algorithm available here: \url{https://github.com/teamRIM/tutoRIM}. \cite{mangeat_super-resolved_2021} 